# DRDST: Low-latency DAG Consensus through Robust Dynamic Sharding and Tree-broadcasting for IoV

Runhua Chen, Haoxiang Luo, *Graduate Student Member, IEEE*, Gang Sun, *Senior Member, IEEE*, Hongfang Yu, *Senior Member, IEEE*, Dusit Niyato, *Fellow, IEEE*, and Schahram Dustdar, *Fellow, IEEE*

*Abstract*—The Internet of Vehicles (IoV) is emerging as a pivotal technology for enhancing traffic management and safety. Its rapid development demands solutions for enhanced communication efficiency and reduced latency. However, traditional centralized networks struggle to meet these demands, prompting the exploration of decentralized solutions such as blockchain. Addressing blockchain's scalability challenges posed by the growing number of nodes and transactions calls for innovative solutions, among which sharding stands out as a pivotal approach to significantly enhance blockchain throughput. However, existing schemes still face challenges related to *a*) the impact of vehicle mobility on blockchain consensus, especially for cross-shard transaction; and *b*) the strict requirements of low latency consensus in a highly dynamic network. In this paper, we propose a DAG (Directed Acyclic Graph) consensus leveraging Robust Dynamic Sharding and Tree-broadcasting (DRDST) to address these challenges. Specifically, we first develop a standard for evaluating the network stability of nodes, combined with the nodes' trust values, to propose a novel robust sharding model that is solved through the design of the Genetic Sharding Algorithm (GSA). Then, we optimize the broadcast latency of the whole sharded network by improving the tree-broadcasting to minimize the maximum broadcast latency within each shard. On this basis, we also design a DAG consensus scheme based on an improved hashgraph protocol, which can efficiently handle cross-shard transactions. Finally, the simulation proves the proposed scheme is superior to the comparison schemes in latency, throughput, consensus success rate, and node traffic load.

This work was supported in part by the National Key Research and Development Program of China under Grant 2023YFB2704200, and in part by the Natural Science Foundation of Sichuan Province under Grant 2022NSFSC0913. *(Corresponding author: Haoxiang Luo and Gang Sun.)*

R. Chen is with the Glasgow College, University of Electronic Science and Technology of China, Chengdu 611731, China, and also with the James Watt School of Engineering, University of Glasgow, G12 8QQ Glasgow, U.K. (e-mail: 2839936C@student.gla.ac.uk).

H. Luo, G. Sun, and H. Yu are with the Key Laboratory of Optical Fiber Sensing and Communications (Ministry of Education), University of Electronic Science and Technology of China, Chengdu 611731, China (email: lhx991115@163.com; {gangsun, yuhf} @uestc.edu.cn).

D. Niyato is with the College of Computing and Data Science, Nanyang Technological University, Singapore 639798 (e-mail: dniyato@ntu.edu.sg).

S. Dustdar is with the Distributed Systems Group, TU Wien, Vienna 1040, Austria, and also with the ICREA, Universitat Pompeu Fabra, Barcelona 08002, Spain. (e-mail: dustdar@dsg.tuwien.ac.at).

## I. INTRODUCTION

THE advent of the Internet of Vehicles (IoV) marks a significant milestone in the evolution of intelligent transportation systems. As a subset of the Internet of Things (IoT), IoV facilitates the exchange of data between vehicles and with infrastructure, enhancing traffic flow, safety, and user experience [1], [2]. To address the security needs of IoV, blockchain offers a promising avenue by enabling decentralized data storage and cryptographic assurance of data integrity [3], [4]. However, the high-speed mobility and large-scale interaction of vehicle communication put forward extremely strict requirements on the rate of information exchange between vehicles. Traditional blockchains have shortcomings in terms of delay and scalability due to complex consensus, broadcast processes.

To surmount these obstacles, sharding has emerged as a feasible solution. By partitioning the blockchain network into smaller, manageable segments, or "shards," it becomes possible to process transactions and achieve consensus in parallel, thereby enhancing throughput and reducing latency [5]. Initially, sharding schemes were predominantly based on random node allocation, such as Elastico [6], OmniLedger [7], and RapidChain [8]. However, random sharding schemes can compromise the security of blockchain systems, as a disproportionate number of Byzantine nodes may be aggregated within one shard, causing 51% of attacks. Therefore, the subsequent focus shifted towards incorporating trust mechanisms, such as Trust-Based Shard Distribution (TBSD) [9], and the methods proposed in [10] and [11]. This method can balance the concentration of malicious nodes within each shard.

Additionally, the IoV involves a substantial amount of data exchange between vehicles and infrastructure, implying that the sharding system must achieve high throughput to efficiently manage traffic information. Hence, researchers have focused on combining sharding with DAG architecture in recent years. The sharding-hashgraph framework can provide a high-performance solution for blockchain [12], even with a low scalable Practical Byzantine Fault Tolerance (PBFT) consensus, which can greatly improve transaction throughput and scalability [13].

## A. Research Motivation

Although the introduction of DAG can lead to a blockchain network with higher throughput, achieving rapid ordering of transactions remains the most significant challenge [14], [15], [16], especially under high-concurrency conditions. The necessity of transaction ordering in blockchain systems is crucial for supporting smart contracts, which rely on a consistent and irreversible sequence of events to execute contract terms autonomously and securely [15], [16]. This implies that in IoV, cross-shard communication resulting from the mobility of vehicles and transaction processing in such dynamic environments require more effective mechanisms to ensure data consistency and order.

Fortunately, tree-broadcasting [17], [18], [19], [20], as a structured protocol, establishes strict parent-child communication relationships, reducing unnecessary message replication and transmission, which means that we can leverage it as an important bridge between sharding and DAG, ensuring the quality of throughput with efficient communication. Additionally, to construct a stable tree-broadcasting structure among nodes, robust sharding must be performed first, as sharding determines with whom to communicate, while the tree structure determines the mode. We find that the two are sequentially dependent. However, the unpredictable behavior of nodes implies that sharding should be dynamic, capable of adjusting the nodes divided into each shard in response to changes in the network environment, thereby ensuring the robustness of the sharding process. This is all in line with our key motivation: to meet the low-latency demands of IoV while achieving high throughput and a high consensus success rate.

## B. Our Contributions

To address the shortcomings of existing work and the above findings, we propose DRDST, a novel low-latency DAG consensus scheme tailored for IoV. The contributions of this paper are as follows

- We introduce a network stability scoring criterion based on node continuous online time, computing power, and failure probability. This standard enables the construction of robust and dynamically adaptive shard models.
- In order to solve a complex multi-objective optimization problem generated by the shard model, we design a new Genetic Sharding Algorithm (GSA), determine the most reasonable sharding scheme, and ensure a secure and more efficient consensus process.
- To improve latency, we introduce the S-MLBT (S-Minimum Latency Broadcast Tree, where $S$ denotes a set of $q$ shards), denoting our tree-based broadcasting. This strategy ensures that information propagates efficiently within each shard, significantly reducing the maximum broadcast latency among shards.
- Shifting from a gossip-based to a tree-based communication pattern, we improve traditional hashgraph protocol. Our DAG consensus mechanism efficiently handles cross-shard transactions and events, ensuring the integration and consistency of blockchain, even for vehicle replacement shards.

## C. Organization of the paper

The rest of the paper is arranged as follows. Section II reviews the related work. Section III delves into the system model, providing a comprehensive framework of our IoV network and DRDST architecture. In Section IV, we provide the details of the robust sharding model. In Section V, we introduce the S-MLBT, explaining our tree broadcasting protocol. Our DAG consensus scheme is covered in Section VI. Section VII presents the simulation results that evaluate the performance of the proposed DRDST protocol against other advanced schemes. Section VIII concludes this paper.

## II. RELATED WORK

### A. Sharding in IoV

The integration of sharding into the IoV landscape represents a significant stride towards addressing the efficiency and scalability challenges inherent in decentralized vehicular networks. Several pioneering studies have endeavored to harness the potential of sharding in IoV.

Singh *et al.* [10] introduced a decentralized trust management scheme for the IoV that leverages blockchain sharding to reduce the main chain's workload and increase transaction throughput. However, the complexity of smart contracts could potentially impair the system's real-time responsiveness and scalability. Moreover, the scheme may not adequately address the effects of vehicle mobility on blockchain consensus, especially in scenarios where vehicles frequently switch between different RSUs. Zhang *et al.* [13] presented a dynamic network sharding approach to enhance the efficiency of the PBFT consensus algorithm in the IoV by minimizing the number of consensus nodes. This method aims to achieve a more focused and efficient consensus. However, it may not fully address the preservation of consensus stability and reliability amidst frequent network topology changes driven by vehicle mobility. Chen *et al.* [21] introduced VT-chain, a vehicular trust blockchain framework that employs a multi-shard system to tackle scalability and performance issues within vehicular networks. It incorporates a hierarchical Byzantine Fault Tolerance (HierBFT) consensus protocol to maintain efficient consensus across shards. Despite the integration of Trusted Execution Environments (TEEs), the framework's sharding configuration still struggles to effectively accommodate the high mobility of vehicles within the network.

Based on these understandings, our work presents an innovative sharding model designed to overcome the limitations of previous approaches by dynamically adjusting the nodes divided into each shard in response to changes in the network environment and optimizing consensus mechanisms for supporting cross-shard transactions and reduced latency. However, current sharding schemes in IoV have not effectively addressed the challenges posed by the high mobility of vehicles affecting cross-shard transaction consensus and the stringent low latency requirements within highly dynamic networks.

## B. Broadcasting

The foundational work in the domain of broadcasting protocol has inspired further enhancements in the efficiency of broadcasting within IoV. Demers *et al.* [22] introduced gossip, a robust algorithm for replicating database updates across all replicas, but it has high message overhead and propagation delay issues and doesn't handle dynamic network topologies or efficient broadcasting well. Rohrer *et al.* [17] introduced the Kadcast protocol, which improves the traditional gossip-based broadcast mechanism by employing a structured tree approach derived from the Kademlia protocol [23], later enhanced with congestion control in Kadcast-NG [18]. However, Kadcast doesn't account for network heterogeneity and node dynamics. Zhu *et al.*'s MLBT [19] aimed to reduce latency and enhance stability in blockchain systems by leveraging network heterogeneity, but it lacks scalability and adaptability for highly dynamic networks. Zheng *et al.*'s DHBN [20] incorporates MLBT to address the challenges of network dynamics and heterogeneity in blockchain networks by maintaining multiple MLBTs and balancing relay tasks among participants.

In the realm of IoV, multi-hop broadcast is vital for timely data dissemination. Selecting the next hop in multi-hop routing is crucial for performance and application reliability. However, high mobility, dynamic network topologies, and complex channel conditions pose challenges to optimal next-hop selection. Palazzi *et al.* [24] proposed a fast multi-hop broadcast algorithm that may struggle with GPS limitations in areas with poor coverage or obstructions. Tian *et al.* [25] introduced a protocol based on geographic coordinates for Vehicular Ad Hoc Network (VANET), potentially ineffective in complex traffic scenarios. Wang *et al.* [26] developed a crowdsensing model with a cluster-optimized framework and a delay-sensitive routing algorithm for IoV event propagation, using stochastic theory and intersection data for relay selection, but it could be vulnerable to misinformation, affecting vehicle and traffic management.

The insights gleaned from the aforementioned studies drive our exploration into the feasibility and effectiveness of employing tree broadcasting in blockchain-based IoV (BIoV). Our focus is on harnessing the protocol's inherent efficiency in reducing communication latency and overhead, and tailoring it to the specific demands of IoV environments. However, existing broadcasting schemes in IoV demonstrate limitations in addressing the dynamic network topologies, high mobility, and complex channel conditions, which collectively hinder their ability to ensure reliable, efficient, and timely message dissemination in diverse traffic scenarios.

## C. DAG Consensus

In pursuing enhancing the throughput of blockchain and scalability, numerous studies have delved into integrating DAG with blockchain systems [27]. The GHOST protocol [28] deviates from the longest-chain rule by selecting the heaviest subtree in blockchain forks, enhancing block validation but risking centralization. Conflux [14] uses a tree graph with parent and reference edges, complicating system implement-tation. IOTA's Tangle [29] replaces blocks with a DAG, eliminating mining and fees, but faces scalability and security issues, particularly Sybil attacks. SPECTRE [15] focuses on honest transaction sequencing through a voting algorithm within a DAG, but its complexity and computational demands are high. PHANTOM [16] builds on SPECTRE, creating a scalable consensus protocol that establishes a total order across a DAG of blocks (blockDAG), distinguishing between compliant and intransigent nodes.

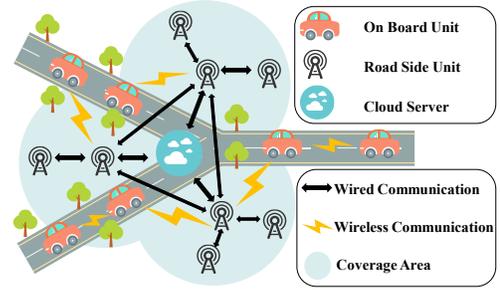

**Fig. 1.** The IoV model.

In response to the burgeoning development of DAG technology, scholars have been proactively integrating DAG structures into the IoV systems to enhance the reliability and efficiency of data sharing. Lu *et al.* [30] introduce a hybrid blockchain architecture known as PermiDAG, which combines a permissioned blockchain with a local DAG to enhance data security and reliability in the IoV. Cui *et al.* [31] presented a Blockchain-Based Containerized Edge Computing Platform (CUTE) designed for the IoV. The platform schedules DAG-based computation tasks efficiently and integrates blockchain technology to enhance network security and reduce computation latency in IoV applications. Fu *et al.* [32] proposed a DAG blockchain-based framework that incorporates a two-stage Stackelberg game for optimizing bandwidth allocation and pricing strategies in the IoV.

However, the common limitation in existing DAG-based blockchain systems is the challenge of achieving rapid ordering of the transactions, particularly under high-concurrency conditions prevalent in IoV. Besides, cross-shard events are not efficiently handled. Our proposed scheme addresses these issues by integrating robust dynamic sharding with tree-broadcasting to optimize transaction processing and order, thereby enhancing the efficiency and security of consensus within the IoV network.

## III. SYSTEM MODEL

### A. Network Model

At the vanguard of vehicular communication systems, IoV integrates advanced sensors, powerful processors, and resilient communication protocols, promising significant improvements in traffic safety, efficiency, and connectivity [33]. As shown in Fig. 1, our network model is designed with a focus on three critical components: the On-Board Unit (OBU), the Roadside Unit (RSU), and the Cloud Server (CS), each playing a distinct role in the ecosystem.

**OBU:** The OBUs are the mobile traffic data collectors with a dynamic topology that enables data aggregation from diverse sources. Despite their limited computational and storage capabilities and the instability of network connections due to mobility, they actively send the collected traffic data to each other and to the RSUs through wireless communication. Since they frequently switch between different RSUs, cross-shard transactions can be created.

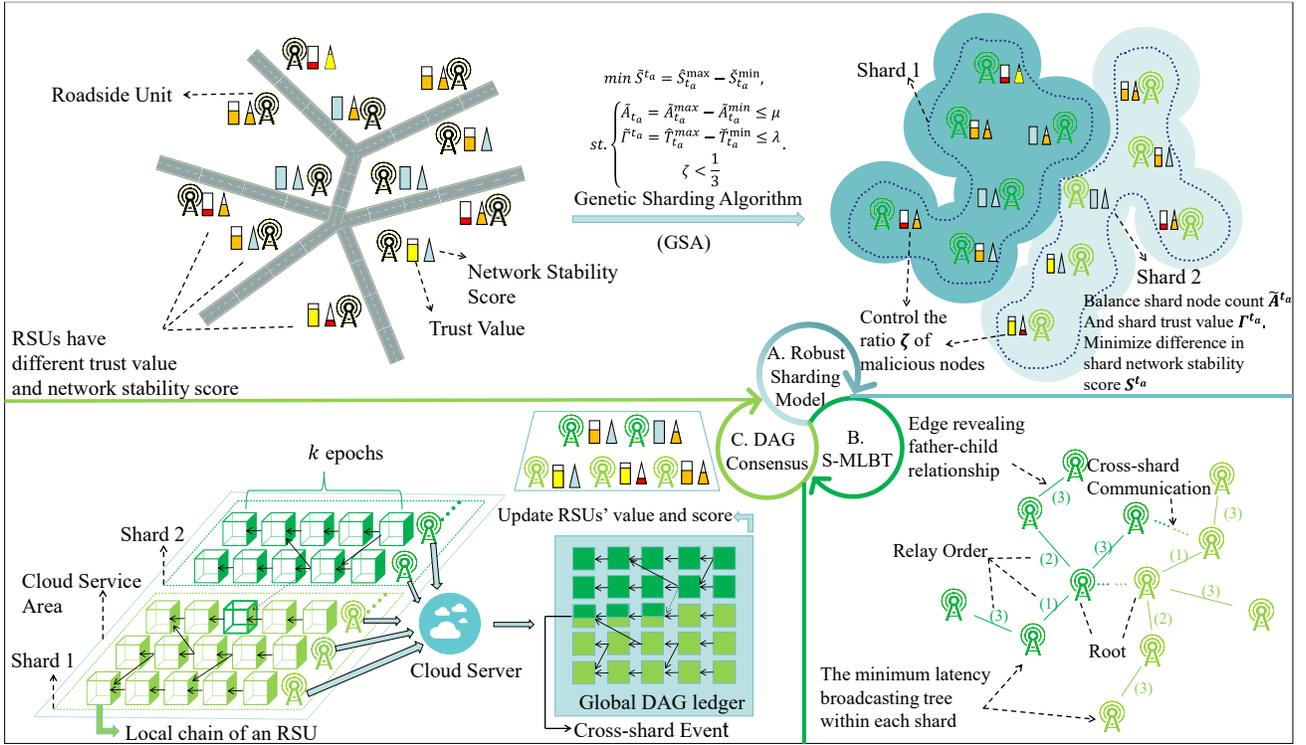

**Fig. 2.** The overall architecture of DRDST.

**RSU:** The RSUs serve as the network's backbone, executing consensus algorithms due to their stationary nature, providing reliable connections, greater bandwidth, storage, and computational power compared to OBUs. The RSUs receive data from OBUs, utilize a tree-based communication pattern for consensus, and record outcomes on local ledgers before syncing with the Cloud Server.

**CS:** The CS acts as the orchestrator, possessing substantial computational and storage capabilities. It manages the RSU sharding algorithm, maintains the global DAG ledger, calculates sharding results for each round, and aggregates consensus outcomes from RSUs across shards. The centralized ledger storage provides a comprehensive transaction record and facilitates the swift integration of new, trustworthy RSUs by enabling them to download the latest ledger and engage in consensus activities immediately.

Similar to [10] and [21], our strategy implements sharding at the RSU layer rather than at the vehicle layer. This approach leverages the fact that RSUs provide more stable connectivity and have superior computational capabilities than OBUs. By sharding RSUs, the network addresses the challenges posed by the mobility of vehicles, able to efficiently handle a growing number of transactions and nodes, thus enhancing the performance and scalability. Our approach takes full account of nodes' potential behavior and conducts a more comprehensive evaluation of each RSU before implementing sharding.

TABLE I
KEY NOTATIONS

| Notations | Definitions |
|---|---|
| $X, Y$ | Set of nodes and set of shards, respectively |
| $\tau_i^{t_a}, s_i^{t_a}$ | Trust value and network stability score of node $x_i$ in time $t_y$, respectively |
| $\Gamma_j^{t_a}, S_j^{t_a}$ | Trust value and the network stability score of shard $j$, respectively |
| $\theta(t_i^o)$ | Trust reduction of node $i$ communicating with node role $o$ |
| $\hat{k}_i^{t_a}, \check{k}_i^{t_a}$ | Number of successfully and unsuccessfully processed transactions by nodes $i$ in term $t_y$, respectively |
| $t_{i,t_a}^o, T_{t_a}^o$ | Actual and required processing time by node $i$ to validate and package transactions into a micro block, respectively |
| $T_{online}$ | Continuous online time of a node |
| $C, \eta$ | Computational capacity and failure probability of a node |
| $\tilde{A}_{t_a}^{max}, \tilde{A}_{t_a}^{min}$ | The maximum and minimum number of RSUs among $s$ shards, respectively |
| $\hat{T}_{t_a}^{max}, \check{T}_{t_a}^{max}$ | The highest and lowest trust value among all the shards, respectively |
| $\hat{S}_{t_a}^{max}, \check{S}_{t_a}^{min}$ | The highest and lowest network stability scores among all the shards, respectively |
| $\tilde{A}_{t_a}, \tilde{\Gamma}^{t_a}, \tilde{S}^{t_a}$ | The difference in node counts, trust value, and network stability score among all the shards, respectively |
| $P_a, P_b, P_c$ | The three candidate solutions selected from the population |
| $M[j], T[j]$ | The mutated and trial individual, respectively |
| $R_{(m,n)}, d_{(m,n)}$ | The data transmission rate and the physical distance between RSU $m$ and $n$, respectively |

*B. DRDST Architecture*

Our proposed DRDST scheme has three major components, the robust sharding model, the S-MLBT broadcasting, and the DAG consensus. Fig. 2 depicts the overall architecture diagram of DRDST. The workflow starts from the initialization of RSUs distributed in different areas. Each RSU is annotated with its trust value and network stability score, which are critical metrics for the sharding algorithm. Upon the cloud server's execution of the algorithm named GSA, which is a more adopted method rather than random [6], [7], [8], RSUs are dynamically assigned to different robust shards. Within each shard, RSUs, guided by the establishment of S-MLBT, engage in a consensus process using an improved hashgraph algorithm. This process results in the formation of local blockchains by each RSU, which are then integrated into a global DAG ledger stored on the cloud server. The DAG structure ensures a coherent and secure ledger across the network, providing a comprehensive view of all transactions and enhancing the overall reliability of the IoV. The design details will be provided from Section IV to Section V. The key notations used in our design are succinctly defined in Table I.

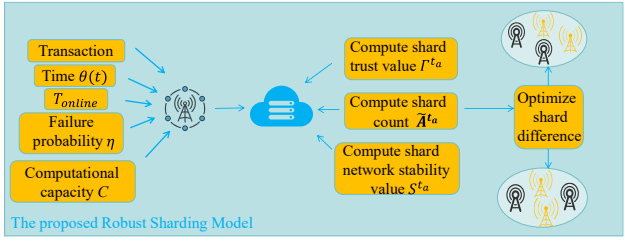

**Fig. 3.** The workflow of our robust sharding model.

## IV. ROBUST SHARDING MODEL

### A. Node Trust and Stability

Network dynamics refers to the frequent changes in the state of nodes within a network, such as the alterations in network topology caused by the movement of vehicles [34]. Conversely, network stability denotes the ability of a network to maintain consistency and reliability in the face of dynamic changes [1], [35]. In the design of our sharding model, to ensure the robustness and reliability of the network, we have formed a new scoring criterion for the trust value and network stability score of nodes as key evaluation metrics. The trust value reflects the behavioral performance of nodes during consensus formation, while the network stability score provides a comprehensive assessment of node performance in the network environment.

The workflow of our model is illustrated in Fig. 3. Initially, the model calculates three critical metrics: node count, shard trust value, and shard network stability score. Subsequently, the model determines the differences in these metrics among all the shards. Finally, some related optimization objectives are utilized to guide the sharding algorithm. The detailed concepts and design of the model are provided as follows.

Let $X$ denote the set of $p$ nodes[1] in a blockchain system

$$X = \{x_i | i \in \{1,2,\ldots,p\}\}, \tag{1}$$

where a node can transmit and validate transactions.

In a time range with a series of time slots $(t_0,\ldots,t_a,\ldots)$ node $x_i$ in time $t_a$ has the form

$$x_i = <Id, \tau_i^{t_a}, s_i^{t_a}>, \tag{2}$$

where $Id$, $\tau_i^{t_a}$ and $s_i^{t_a}$ are the node identification number, trust value, network stability score of node $x_i$, respectively.

Considering that each node will communicate with three types of nodes, namely Root (Ro), Father (Fa), and Child (Ch), according to the tree-broadcasting protocol which would be later introduced in Section IV, we compute $\tau_i^{t_a}$ as follows:

$$\tau_i^{t_a} = \alpha(\hat{k}_i^{t_a} - \check{k}_i^{t_a}) - \beta\left(\frac{\theta(t_i^{Ro})}{T_{t_a}^R} + \frac{\theta(t_i^{Fa})}{T_{t_a}^F} + \frac{\theta(t_i^{Ch})}{T_{t_a}^C}\right), \tag{3}$$

$$\theta(t_{i,t_a}^o) = \begin{cases} 0, & t_{i,t_a}^o \leq T_{t_a}^o, o\epsilon\{Ro, Fa, Ch\} \\ t_{i,t_a}^o, & otherwise \end{cases}, \tag{4}$$

where $\alpha$ and $\beta$ are the trust reward and reduction coefficients, respectively; and $\theta(t_i^o)$ denotes the trust reduction of node $i$ communicating with node role $o$. $\hat{k}_i^{t_a}$ and $\check{k}_i^{t_a}$ are the number of successfully and unsuccessfully processed transactions by nodes $i$ in term $t_a$. $t_{i,t_a}^o$ and $T_{t_a}^o$ are the actual and required processing time by node $i$ to generate a micro block in $t_a$, respectively. Required time is determined by the average processing time across all RSUs for generating a block.

[1] The nodes involved in our model are all RSUs.

The main idea behind this design is that reaching consensus should be rewarded, while stalling time should be punished. Those who mine selfishly and do not want to forward the message, or directly forward the tampered message will all suffer from a decrease in trust value. The network stability score is directly related to the reliability and efficiency of nodes, data transmission, and verification in the sharded network. A node with poor network stability has a shorter continuous online time $T_{online}$, a lower computational capacity $C$ and a higher failure probability $\eta$. We compute $s_i^{t_a}$ by

$$s_i^{t_a} = \sum_{k=1}^{b} w_k f_k(\Omega_k^{t_a}), \tag{5}$$

where $w_k > 0$ is the weight coefficient determined through Analytic Hierarchy Process (AHP) [36], corresponding to the $b$ indicators, namely $T_{online}$, $C$, and $\eta$. $\Omega_k^{t_a}$ represents the value of the $k$-th indicator at time $t_a$, and $f_k(\Omega_k^{t_a})$ is the nonlinear function that maps $\Omega_k^{t_a}$ to the score range of [0,1]. We show $f_k$ functions for different values of $k$ in next three equations.

The score judging from $T_{online}$ is calculated through the Sigmoid function, which provides a smooth output, mapping linear online data to non-linear values in the interval [0, 1].

$$f_1(T_{online}) = \frac{1}{1 + e^{-(T_{online} - T_0)}}, \tag{6}$$

where $T_0$ is the average online time.

When an RSU joins the communication tree, it and its connected nodes send an online message to the CS with the join timestamp. Upon disconnection, they notify the CS with an offline timestamp. With these timestamps, the CS calculates each node's online duration and the overall $T_0$ across the network. Additionally, when an RSU leaves, the CS resets its $T_{online}$ to zero.

From a statistical point of view, since RSU devices may be due to differences in cost, design, or functional requirements, the distribution of computing resources $c$ is uneven, subject to a positive skew distribution. We use logarithmic functions to process the data, reduce the influence of extreme values in the distribution, and use min-max normalization to calculate the score judging from $C$.

$$f_2(C) = \frac{log(C) - min(log(C))}{max(log(C)) - min(log(C))}. \quad (7)$$

Assuming in the IoV, the probability of each RSU being attacked is independent of time, that is, each attack is an independent event [35]. In our standard, if the RSU has never been paralyzed ($\eta = 0$), the score is 1, indicating that the node is very stable. As failure probability $\eta$ increases, the score gradually decreases, approaching 0. This feature can be characterized by an exponential decay function as follows:

$$f_3(\eta) = e^{-\gamma \eta}, \quad (8)$$

where $\eta$ represents the failure probability of the RSU, and $\gamma$ is a positive adjustment parameter used to control the rate of decline of the scoring function.

This flexibility is essential in a network where conditions can change rapidly. For instance, a higher $\gamma$ would lead to a faster decrease in score as the failure probability $\eta$ increases, which could be desirable in environments where quick detection and response to potential security threats are critical.

### B. Constrained Optimization Problem

Firstly, when sharding, the number of nodes within each shard must be taken into account. If two shards have the same shard trustworthiness values but different numbers of nodes, the shard with more nodes might have lower individual node trustworthiness. Let $\tilde{A}_{t_a}$ represent the difference in node counts among shards as follows:

$$\tilde{A}_{t_a} = \tilde{A}_{t_a}^{max} - \tilde{A}_{t_a}^{min}, \quad (9)$$

where $\tilde{A}_{t_a}^{max}$ and $\tilde{A}_{t_a}^{min}$ are the maximum and minimum RSU counts among $q$ shards. By minimizing $\tilde{A}_{t_a}$, we strive to achieve balanced consensus speeds across all shards.

Secondly, to prevent the trust value and network stability of some shards too low, we need to allocate nodes to shards as reasonably as possible. Otherwise, the shard may be a single shard taken over by malicious nodes, resulting in poor reliability and security of the entire blockchain sharding system. Let the trust value and network stability score of a shard be represented by the average value of all the nodes in the shard, then, we have

$$\Gamma_j^{t_a} = \frac{\sum_{x_i \in \delta_j} \tau_i^{t_a}}{|\delta_j|}, \quad (10)$$

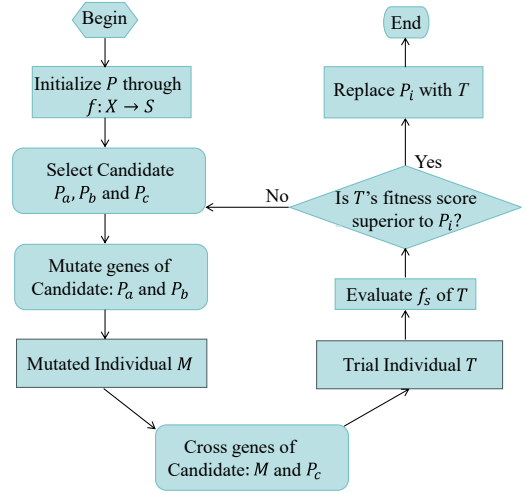

**Fig. 4.** The workflow of Genetic Sharding Algorithm.

$$S_j^{t_a} = \frac{\sum_{x_i \in \delta_j} s_i^{t_a}}{|\delta_j|}, \quad (11)$$

where $\Gamma_j^{t_a}$ is the trust value of shard $j$, $\delta_j$ is the set of nodes in shard $j$ and $S_j^{t_a}$ is the network stability score of shard $j$. Therefore, we have

$$\tilde{\Gamma}^{t_a} = \hat{T}_{t_a}^{max} - \check{T}_{t_a}^{min}, \quad (12)$$

$$\tilde{S}^{t_a} = \hat{S}_{t_a}^{max} - \check{S}_{t_a}^{min}, \quad (13)$$

where $\hat{T}_{t_a}^{max}$ and $\check{T}_{t_a}^{min}$ are the highest and lowest trust values in $q$ shards, respectively; $\hat{S}_{t_a}^{max}$ and $\check{S}_{t_a}^{min}$ are the highest and lowest network stability score in $q$ shards, respectively. Through minimizing $\tilde{\Gamma}^{t_a}$ and $\tilde{S}^{t_a}$, we prevent security bottlenecks in the sharding system.

Since robustness is crucial for blockchain systems, we choose $\tilde{S}^{t_a}$ as the primary optimization target, while converting $\tilde{\Gamma}^{t_a}$ and $\tilde{A}_{t_a}$ into constraints.

$$min \; \tilde{S}^{t_a} = \hat{S}_{t_a}^{max} - \check{S}_{t_a}^{min},$$

$$st. \begin{cases} \tilde{A}_{t_a} = \tilde{A}_{t_a}^{max} - \tilde{A}_{t_a}^{min} \leq \mu \\ \tilde{\Gamma}^{t_a} = \hat{T}_{t_a}^{max} - \check{T}_{t_a}^{min} \leq \lambda \\ \zeta < \frac{1}{3} \end{cases}, \quad (14)$$

where $\mu$ represents the acceptable difference in RSU shard node counts, $\lambda$ represents the trust value threshold, and $\zeta$ represents the ratio of malicious nodes in the shard.

### C. Design of Genetic Sharding Algorithm

Considering the vast number of nodes and the diversity of trust values and network stability scores among nodes, it is imperative to employ an algorithm capable of handling large-scale, multi-objective optimization challenges. The Genetic Algorithm (GA) [37], a heuristic search algorithm that simu-

**Algorithm 1:** Fitness Function

**Input:** Set of nodes $X$, shard assignment $T$
**Output:** Fitness score $f_s$
1    $S \leftarrow$ default dict (list)
2    **Step1: Assign all nodes to their corresponding shards and form all the shards.**
3    **for** each $(i, \text{shard})$ in enumerate $(T)$ **do**
4      $S[\text{shard}] \leftarrow x_i$
5    **end for**
6    **Step2: Calculate trust value, network stability score and node count of each shard.**
7    **for** $s_j$ in $S$ **do**
8      Calculate $\Gamma_j^{ta}$ using Eq. (10)
9      Calculate $S_j^{ta}$ using Eq. (11)
10     $\widetilde{A}_i^{ty} \leftarrow |y_j|$
11    **end for**
12    **Step3: Calculate the difference of the three metrics among all the shards.**
13    Calculate $\widetilde{\Gamma}^{ta}$ using Eq. (12)
14    Calculate $\widetilde{S}^{ta}$ using Eq. (13)
15    Calculate $\widetilde{A}_{t_a}$ using Eq. (9)
16    **Step4: Apply penalty value to judge whether any shard exceeds predefined threshold on these metrics.**
17    penalty $\psi \leftarrow 0$
18    **if** $\widetilde{\Gamma}^{ta} > \lambda$ or $\widetilde{A}_{t_a} > \mu$ or $\zeta > \frac{1}{3}$ or $\widetilde{S}^{ta} > 0.15$ **then**
19      $\psi \leftarrow \psi + 1000$
20      $f_s \leftarrow \Gamma_j^{ta} + \widetilde{A}_{t_a} + \widetilde{S}^{ta} + \zeta + \psi$
21    **end if**
22    **return** $f_s$

lates the process of biological evolution, stands out for its global search capability and robustness against variations in problem scale. Since fast response to a highly dynamic network is especially significant in IoV, we form our GSA to enable rapid and accurate sharding.

As depicted in Fig. 4, the workflow of the GSA, inspired by genetic principles, is designed to accommodate RSU sharding system and encompasses several critical steps.

*1) Initialization of the Population:* Let $P = \{P_1, P_2, \ldots, P_{\text{population size}}\}$ be the current population, where $P_u$ represents a candidate solution, i.e., the mapping scheme from RSUs to shards. The initial candidate solution population is generated through a randomization mechanism, ensuring the diversity of the population and providing a broad exploration space for the evolution of the algorithm. Specifically, for an individual $P_u$ in the population, where $u$ indicates the index in the population. Each element $P_u[i]$ for nodes, where $i$ is the index of the node, is randomly selected from the network. This initialization strategy not only ensures the diversity of the population but also provides a broad exploration space for the evolution of the algorithm. Then, we can express the initialization process as a mapping relationship from the node set $X$ to the shard set $Y$. Let $Y = \{y_j | j \in \{1,2,\ldots,q\}\}$. be the shard set, and the mapping function for initializing the population $f: X \rightarrow Y$ can be represented as:

$$f(x_i) = y_j. \quad (15)$$

*2) Iteration:* The iterative process is the driving force behind genetic algorithms, which continuously optimizes the population through selection, crossover, and mutation operations. The iterative process can be described as follows.

- **Selection operation:** Randomly select three different candidate solutions $P_a$, $P_b$ and $P_c$ from the current population $P$ as samples. This can ensure each candidate solution has an equal probability be selected.
- **Mutation operation:** We denote the network stability score of an RSU at gene position (located shard) $j$ in $P_a$ as $s_a[j]$ and in $P_b$ as $s_b[j]$. With the Mutation Factor (MF) as the condition, choose genes from $P_b$ to replace corresponding positions in $P_a$, generating a mutated individual $M$.

$$M[j] = \begin{cases} P_b[j] & \text{if } rand(0,1) < MF\left(1 - \frac{s_a[j] + s_b[j]}{2s_{max}}\right) \\ P_a[j] & \text{otherwise} \end{cases}, \quad (16)$$

where $s_{max}$ is the maximum network stability score among all RSUs. The MF is here modified to be inversely proportional to the network stability score, encouraging the retention of more stable RSUs.

- **Crossover operation:** Let $s_M[j]$ and $s_{P_c}[j]$ denote the network stability scores of the RSUs at position $j$ in $M$ and $P_c$, respectively. With the crossover probability (CP) as the condition, choose genes from the mutated the individual $M$ to replace corresponding positions in $P_c$, generating a trial individual $T$.

$$T[j] = \begin{cases} M[j] & \text{if } rand(0,1) < CP\left(1 - \frac{s_M[j] + s_{P_c}[j]}{2s_{max}}\right) \\ P_c[j] & \text{otherwise} \end{cases}. \quad (17)$$

This modified CP ensures that RSUs with higher network stability are more likely to be selected for the new sharding configuration.

*3) Fitness evaluation:* It is crucial for evaluating the quality of candidate solutions. In GSA, the fitness function takes into account the trust value of nodes within each shard $\widetilde{T}_j^{ta}$, network stability score $S_j^{ta}$, and the size of the shard $\widetilde{A}_i^{ta}$. If any shard exceeds the predefined threshold on these metrics, a penalty term $\Psi$ will be applied to the fitness score $f_s$ of the trial individual, guiding the algorithm toward better shard structures. This process is summarized in Algorithm 1.

*4) Selection operation:* If $T$'s fitness is superior to $P_u$, then use $T$ to replace $P_u$ to form a new population $P'$. The next iteration will operate on this new population.

Since the trust value and network stability score of each node are updated regularly, sharding among RSUs is dynamic and the influence of malicious nodes on the consensus results can be reduced, thus alleviating the challenges brought by the Byzantine problem to a certain extent.

*D. Complexity Analysis*

We consider both the time complexity and the space complexity of the GSA. The time complexity primarily focuses on the time required for the execution of the algorithm, while the space complexity is concerned with the storage space needed during the execution process. $U$ is defined as the population size, which is the total number of individuals considered in each generation; $p$ as the total number of nodes, representing the complexity of the problem space; and $I$ as the length of each individual's genome, indicating the number of genes or parameters to be optimized.

*1) Time Complexity Analysis:* The algorithm initiates with a random selection of three unique individuals from the population, incurring $O(N)$ complexity due to the population-wide iteration to prevent duplicates. Mutation follows, comparing genes of two individuals per genome, resulting in $O(I)$ complexity as each gene must be examined. Crossover mirrors this with $O(I)$ complexity, iterating over genes to form trial individuals. Fitness evaluation incurs $O(p)$ complexity, requiring a scan of all nodes and shards. The final selection and replacement operation, assessing and potentially updating individuals in the population, carries $O(Np)$ complexity. Collectively, these operations yield a total time complexity of $O(GNp)$ for the GSA, with $G$ denoting generations, highlighting the algorithm's scalability and computational intensity across iterations.

*2) Space Complexity Analysis:* GSA requires storage for the population and any auxiliary structures used during the fitness evaluation. The population storage alone demands $O(NI)$ space, as each individual in the population must be represented along with their corresponding genes. The new population generated in each iteration also requires the same amount of space, leading to a combined space complexity for population storage of $O(NI)$. Additionally, the auxiliary space for calculating trust scores, stability scores, and shard sizes during the fitness evaluation requires $O(p)$ space, as it involves maintaining lists, dictionaries, sets and tuples for all nodes. Consequently, the total space complexity of the GSA is $O(NI + p)$. This analysis underscores the algorithm's spatial efficiency and the resources needed to support its operations, which is essential for practical applications where memory and storage are critical considerations.

*E. Quality of Solutions*

To analyze the quality of solutions produced by the GSA, the convergence trend of the average fitness over generations is employed as a critical performance indicator. This metric evolves towards optimal solutions with each successive gener-ation. In comparison, we compare GSA with Zhang *et al.*'s [38] sharding method based on Differential Evolution (SDE) and traditional GA [37]. Specifically, we execute each algorithm 1000 times, and for each run, the best fitness value at every generation is recorded. We then aggregate the 1000 best values to obtain the mean fitness for each generation.

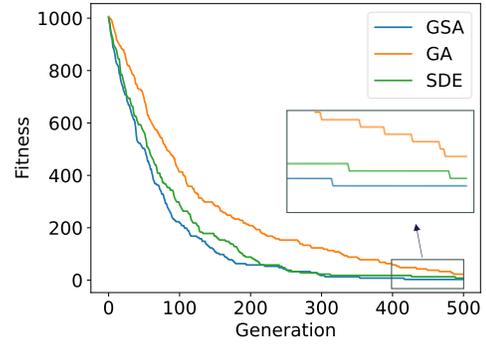

**Fig. 5.** The convergence trend comparison.

The convergence trend of the average fitness over generations, as depicted in Fig. 5, demonstrates effectiveness in identifying and refining high-quality solutions. The GSA exhibits the steepest decline in the early generations, which underscores its most rapid improvement in solution quality during this critical phase. Furthermore, as the algorithms progress into the gradual plateau phase, the GSA maintains the lowest average fitness. This suggests that the GSA is not only efficient in the initial stages but also continues to refine solutions effectively, achieving a near-optimal state in a limited number of generations. This convergence trend is a testament to GSA's potential for real-world applications in optimizing sharding in blockchain-enabled IoV systems.

## V. S-MINIMUM LATENCY BROADCAST TREE

Based on GSA, the RSU network is divided into multiple shards to ensure robustness and low latency of the system.

We construct an MLBT within each shard to further optimize latency, adopting a tree-based communication pattern instead of gossip. To optimize the latency of the entire RSU network, it is ultimately necessary to optimize the MLBT with the maximum latency among the constructed $q$ shards. The problem is formally defined as follows. To build an MLBT for each shard, we need to determine for each root node its set of paths to the other all nodes within the same shard.

The latency is the sum of data transmission latency and propagation latency. The former refers to the latency caused by data transmission over wired channels between microservices. Microservices are the fine-grained, independently deployable service components within the IoV's edge computing environment. The latter refers to the time it takes for a signal to travel through the propagation medium. Thus, the total latency between RSU $m$ and $n$ is

$$t_{wired} = \begin{cases} \dfrac{L}{R_{m,n}} + \dfrac{d_{m,n}}{v}, & m \neq n \\ 0, & otherwise \end{cases}, \quad (18)$$

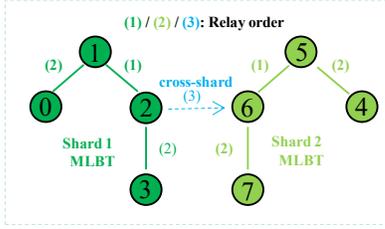

**Fig. 6.** Two MLBTs ($H=2$) from shard 1 and shard 2.

where $L$ is the size of the data packet transmitted, $R_{m,n}$ is the data transmission rate between RSU $m$ and $n$, $d_{m,n}$ is the physical distance between RSU $m$ and $n$, and $v$ is the propagation speed of the signal in the medium. For each shard, let $W_j$ be the set of all paths $p_j^g$ within the shard $j$.

$$W_j = \{p_j^g | g \in \{1, 2, \ldots, h\}\} \quad (19)$$

where $h$ denotes the number of paths connected by the root node $r_j$ in shard $j$. Our goal is to find a set of paths $W_j$ that satisfies

$$W_j = \arg\min_{W_j} \max_{p_j^g \in W_j} \sum_{E \in p_j^g} t_{E\_wired}. \quad (20)$$

where $E$ is an edge representing one hop from one node to another on the path $p_j^g$, and $t_{E\_wired}$ denotes the latency from RSU $m$ to $n$ on edge $E$. Thus, the optimal solution for the broadcast latency of the entire network should satisfy

$$\min\max_{1 \leq i \leq q} \left( \max_{p_j^g \in W_j} \sum_{E \in p_j^g} t_{E\_wired} \right). \quad (21)$$

Although NP-complete problems lack exact polynomial-time solutions [39], they can also be solved by a common heuristic or approximate algorithm. After solving the problem, we can establish MLBT in each shard. For instance, as is shown in Fig. 6, there are two MLBTs with height $H = 2$ originating from shard 1 and shard 2, respectively. RSU 1 and RSU 5 are root nodes for each MLBT. When RSU 2 from shard 1 initiates consensus, it will broadcast the events to its father node RSU 1 and child node RSU 3. When synchronizing an event, RSU 1 will broadcast the event to fan-out node RSU 6 via cross-shard communication.

## VI. DAG CONSENSUS

### A. Improved Hashgraph

Gossip faces high redundancy and difficulty in ensuring consistency when handling cross-shard transactions. Therefore, the traditional gossip-based hashgraph [40] for DAG consensus makes it difficult to achieve efficient transaction ranking. To avoid these problems, we propose an improved hashgraph to enable DAG consensus. It optimizes the efficiency and security of cross-shard transactions, fundamentally addressing the complexities introduced by the mobility of vehicles in the IoV. In the hashgraph architecture, each node maintains a local chain that is part of the global DAG. This DAG is built through inter-node references, establishing parent-child relationships between events. Each event contains two cryptographic hashes: one for its self-parent and another for its other-parent. This interlinked structure forms an immutable chain, where each event is cryptographically bound to others, enhancing the system's resistance to tampering. Hashgraph ensures that all nodes eventually see every transaction through a consensus that maintains the immutability and order of the ledger.

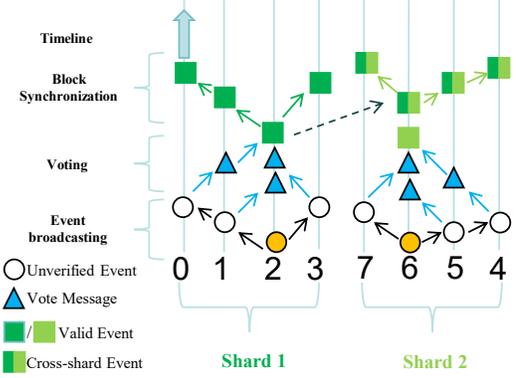

**Fig. 7.** DAG consensus based on improved hashgraph.

In our DRDST, nodes propagate new events via the S-MLBT protocol, which surpasses gossip in network-wide information dissemination. An event achieves consensus and is recorded on an RSU's local chain once it garners over two-thirds YES votes within its shard [40], and is then broadcast to other shards. Utilizing hashgraph, receiving nodes integrate new events with existing transactions and continue broadcasting, ensuring cross-shard transactions are processed and confirmed within the global DAG, as shown in Fig. 7. Transactions will be strongly seen by RSUs across all involved shards, with the hashgraph protocol securing a global order for all events through its DAG structure and consensus. Cryptographic linkage among events deters tampering, ensuring transaction security. Therefore, when vehicles frequently switch between RSUs across different shards, our approach ensures visibility of cross-shard transactions among RSUs, enabling rapid processing of subsequent transactions upon a vehicle's cross-shard movement, thus maintaining efficient, resource-light network operation.

### B. Security Analysis

**Sybil attack:** This attack in distributed systems refers to a malicious strategy where an adversary creates a multitude of pseudonymous identities to gain undue influence over a system that relies on democratic processes or consensus mechanisms. Such attacks threaten the integrity and security of blockchain networks. Traditional hashgraph's resistance to Sybil attacks is derived from its non-permissioned structure, which makes it economically prohibitive for an adversary to control a significant portion of the network with fake identities. Thus, as long as our design combines a secure identity authentication mechanism, such as [1], it can easily resist this attack.

**View split attack:** This attack exploits nodes' limited visibility of transactions to deceive them into validating and spreading fraudulent transactions. Fortunately, our DAG consensus has resistance to view split attacks.

We set each shard to have $u$ RSUs, and $f$ as the maximum number of faulty or Byzantine nodes, where $u \geq 3f + 1$. Consider two conflicting cross-shard transactions involving these $q$ shards, $Tx_1$ and $Tx_2$. Let $U_1$ and $U_2$ represent the sets of RSUs that strongly see these transactions. According to [40], the cardinality of these sets in the sharded network must satisfy

$$|U_1| > 2qf + q,$$
$$|U_2| > 2qf + q. \quad (22)$$

Given the total number of RSUs $qu \geq 3qf + q$, the intersection of $U_1$ and $U_2$ must contain at least $q(f + 1)$ RSUs.

$$|U_1 \cap U_2| \geq q(f + 1), \quad (23)$$

which implies that at least $q(f + 1)$ RSUs can concurrently observe both $Tx_1$ and $Tx_2$. Since the maximum number of malicious nodes in the whole network is $qf$, at least $q$ honest RSUs can accurately discern the legality of the transactions and vote accordingly. An illegal cross-shard transaction will not receive more than $2qf$ votes, thus failing to achieve strong visibility and preventing the view split attack.

## VII. SIMULATION RESULTS

In this section, we compare DRDST with two methods through simulations: the Tree-Based Gossip Protocol (TBGP) [41], and the Graphical Consensus-Based Sharding (GCS) [42].

**TBGP:** It constructs the tree-based broadcasting protocol to optimize the randomness in traditional gossip, thereby reducing redundancy in message dissemination and improving consensus efficiency. It also employs federated learning to select nodes for constructing the TBGN, ensuring network stability and efficiency. Then, it introduces a block weight-based chain selection algorithm to achieve effective global ordering of blocks.

**GCS:** It uses graphical consensus within the shard, forming consensus groups through maximum connected subgraphs and electing leaders based on node reliability. Data within each shard is stored locally on the chain, while the main chain adopts a DAG structure, to support cross-shard sharing. Additionally, the GCS scheme incorporates shard backup and node scheduling strategies to address shard failures and overheating issues.

In our simulation comparison, we extensively evaluate the TBGP, GCS, and DRDST schemes across four metrics.

**Latency ($\vartheta$):**

$$\vartheta = \frac{1}{Z}\sum_{c=1}^{Z}(T_{\text{con},c} - T_{\text{sub},c}). \quad (24)$$

where $Z$ is the total number of transactions, $T_{\text{con},c}$ is the time when the $c$-th transaction reaches consensus and $T_{\text{sub},c}$ is the time when the $c$-th transaction is submitted by the client.

**Throughput ($\Phi$):**

$$\Phi = \frac{\sum_{i=1}^{q} T_{Xi}}{t}, \quad (25)$$

TABLE II
SIMULATIONS PARAMETERS

| Parameters | Values |
|---|---|
| Research area | $100/200\ km^2$ |
| Number of RSUs | 100/200 |
| Number of vehicles | [1000, 10000] |
| Speed of vehicles | $20/40/60/80\ km/h$ |
| Client request speed | 2000/3000 transactions per second |
| Distance between two RSUs | $[500, 1500]\ m$ |
| RSU bandwidth | $[10, 30]\ Mb/s$ |
| Maximum transactions within one event | 1024 |
| Shard count | 4/6/8/10/12 |
| Trust value of each RSU | [0,10] |
| Network stability score of each RSU | [0,1] |
| Number of fan-out nodes of each RSU | 8 |
| RSU Byzantine fault rate | 2%-15% |
| RSU offline/inactive rate | 2%-10% |

where $T_{Xi}$ is the total number of transactions processed by RSU $i$ in the period $t$. The unit for $\Phi$ is transactions per second, $tps$.

**Consensus Success Rate ($\Upsilon$):**

$$\Upsilon = \frac{T_{X_c}}{T_{X_r}}, \quad (26)$$

where $T_{X_c}$ is the number of transactions reached consensus and $T_{X_r}$ is the total number of requested transactions.

**Node Traffic Load ($\omega$):**

$$\omega = \frac{\sum_{i=1}^{q} D_i}{q}, \quad (27)$$

where $D_i$ is the total volume of data transmitted by RSU $i$ in the network. The unit for $\omega$ is MegaByte $MB$.

Through comparative analysis, we demonstrate that the DRDST scheme meets the real-time requirements of vehicular networks while achieving high throughput and consensus success rate, as well as exhibiting efficiency advantages in transaction processing.

### A. Simulation Setup

Our computer is equipped with an Intel i9-14900HX processor, a 2.2GHz CPU, and an NVIDIA GeForce RTX 4070 GPU with 8 GB memory. We implement a simulation of the network and consensus layer for the IoV based on Python 3.0, which accommodates a customizable number of participating clients and nodes. To ensure that our configuration is close to real-world conditions, we incorporate the data transmission rate between RSUs and propagation speed to be the same as [43]. Furthermore, motivated by the need to reflect a realistic deployment scenario, we have considered two areas: one with 100 RSUs and another with 200 RSUs, allowing for a comprehensive evaluation of the network performance. All the parameters are listed in Table II.

### B. Latency

Initially, we set up 100 remote RSUs within a $100\ km^2$ area with a client request speed of 3,000 $tps$. Fig. 8 (a) shows that the latency decreases as the shard count increases. DRDST achieves the lowest latency in all cases by balancing latency among shards and utilizing S-MLBT for efficient communication. TBGP has the highest latency since it focuses more on node stability to ensure robustness. GCS has lower latency than TBGP because it uses asymmetric encryption to protect off-chain data, thus improving computing latency.

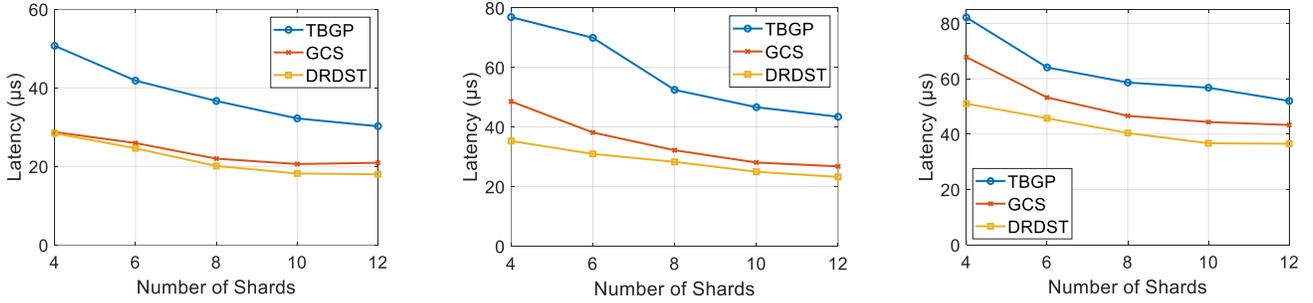

**Fig. 8.** Latency comparison (a) 100 RSUs and 3000 *tps*. (b) 200 RSUs and 3000 *tps*. (c) 200 RSUs and 2000 *tps*.

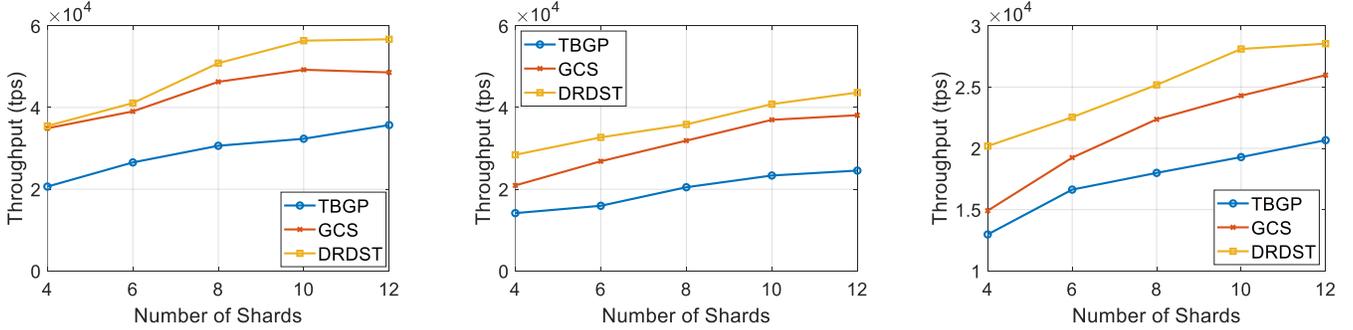

**Fig. 9.** Throughput comparison (a) 100 RSUs and 3000 *tps*. (b) 200 RSUs and 3000 *tps*. (c) 200 RSUs and 2000 *tps*.

Subsequently, we expand the scope to encompass an area with 200 RSUs, while maintaining the same client request speed. This extension enables us to investigate the network's scalability and performance with heightened service unit density. Fig. 8 (b) shows that latency increases for all schemes due to the growing number of nodes in each shard, which demands more extensive communication. DRDST continues to showcase the lowest latency in all cases. However, TBGP's focus on ensuring shard stability and GCS's focus on geographical location may both lead to a single shard becoming a bottleneck.

Ultimately, we decrease the client request speed to 2,000 *tps* and maintained 200 RSUs in the network. As shown in Fig. 8 (c), the latency for all three schemes experiences a further increase under various shard configurations, suggesting that the allocation of resources across shards may not be optimally adapted to the decreased speed. For instance, if certain RSUs become bottlenecks, latency at these points may continue to rise even with an overall reduction in load. Nonetheless, DRDST consistently exhibits the lowest latency in all cases, indicating that the system is capable of reconfiguring resources and adjusting task distribution when shifting from a high-load to a low-load state. This validates the dynamic efficacy of our sharding approach.

### C. Throughput

First, we set up 100 RSUs, with the client request speed set at 3,000 *tps*. Fig. 9 (a) shows that as shard count increases, the parallelism is enhanced, leading to an increase in throughput. DRDST achieves the highest throughput through a balanced sharding strategy and tree-broadcasting protocol. GCS outperforms TBGP by incorporating node scheduling to handle shard overload and expedite the verification of backlog transactions. GCS's lower throughput relative to DRDST is likely due to PBFT's requirement for multiple message exchange rounds, leading to significant communication overhead.

Next, we extend our analysis to include an area with 200 RSUs, while maintaining the same client request speed. Fig. 9 (b) indicates that throughput decreases for all schemes due to the increased node count per shard, which slows transaction propagation and complicates consensus. Despite this, DRDST maintains superior throughput. As shard count increases, GCS's throughput peaks early, TBGP's grows slowly, and DRDST shows consistent improvement. This demonstrates DRDST's superior adaptability to network size changes, optimizing communication routes and resource allocation for sustained high performance.

Finally, we reduced the client request speed to 2,000 *tps* while maintaining a network with 200 RSUs. Fig. 9 (c) reveals a throughput reduction for all schemes, likely due to a mismatch between request speed and network capacity. While higher client request speeds do not guarantee increased throughput, the drop to 2,000 *tps* may significantly underutilize the network's capacity, leading to unrealized throughput gains. Despite this, DRDST consistently outperforms in throughput, highlighting the benefits of dynamic sharding for efficient network resource utilization.

### D. Consensus Success Rate

Commencing our analysis, we set up 100 RSUs under the client request rate of 3,000 *tps*. Fig. 10 (a) shows that the consensus success rate decreases as the shard count increases, as malicious nodes exploit the smaller group to their advantage. DRDST achieves the highest rate and minimal decline, due to sharding through node trust and network stability score. As the shard count rises, GCS suffers the worst decrease in rate, due to its reliance on RSU geography rather than performance metrics. Compared with GCS, TBGP has a higher consensus success rate since it classifies nodes according to their stability.

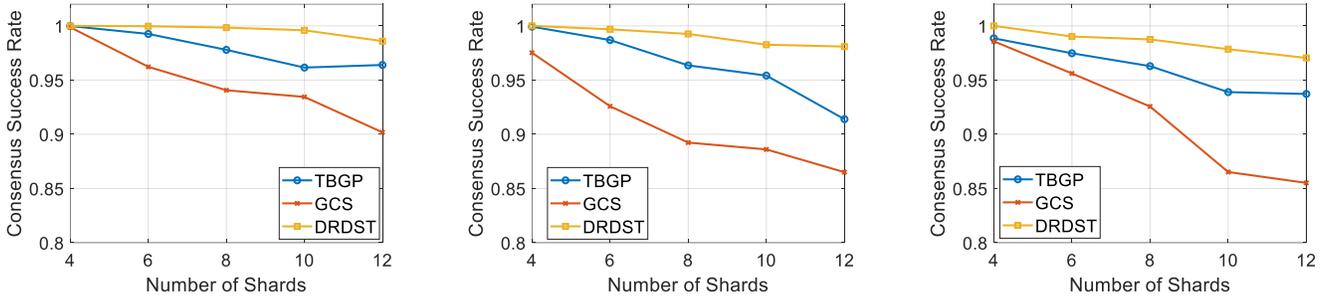

**Fig. 10.** Consensus success rate (a) 100 RSUs and 3000 $tps$. (b) 200 RSUs and 3000 $tps$. (c) 200 RSUs and 2000 $tps$.

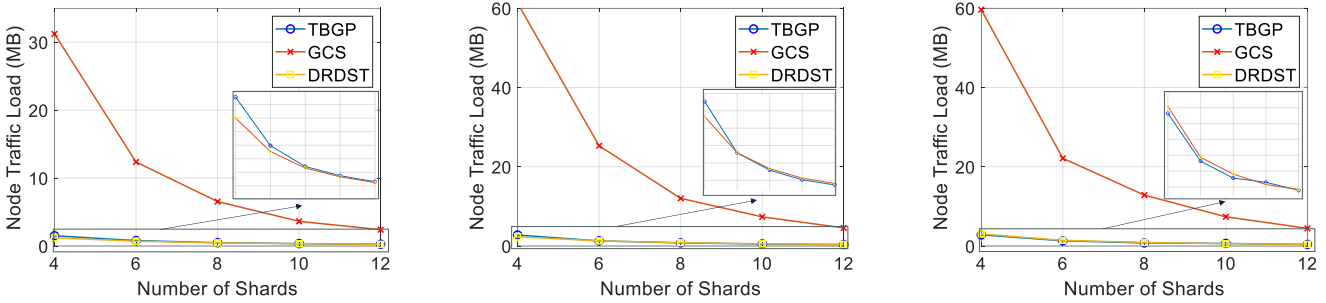

**Fig. 11.** Node traffic load (a) 100 RSUs and 3000 $tps$. (b) 200 RSUs and 3000 $tps$. (c) 200 RSUs and 2000 $tps$.

Following the initial setup, we expand the network to 200 RSUs, persisting with the client request speed at 3,000 $tps$. We elevate the Byzantine fault rate to better assess system behavior under a high fault rate, essential for evaluating network robustness at scale. Fig. 10 (b) shows that the consensus success rate declines for all schemes, reflecting that a higher Byzantine fault rate leads to fewer commit messages or YES votes for consensus initiators. Initially, with 4 shards, all schemes achieve high consensus success rates due to a lower proportion of malicious nodes in each shard. However, as shard count increases, GCS experiences the most significant drop, attributed to its reliance on RSU geography rather than performance, potentially compromising shard integrity. In contrast, TBGP consistently outperforms GCS, as FL categorizes nodes by performance and stability.

Ultimately, we opt to reduce the client request speed to 2,000 $tps$ while maintaining a consistent 200 RSUs. In Fig. 10 (c), DRDST still demonstrates the highest consensus success rate across all shard counts. Besides, as the shard count increases, the DRDST scheme exhibits the smallest decrease at 2.95%, compared to 5.19% for TBGP and 13.25% for GCS, reaffirming the effectiveness of our robust sharding model.

*E. Node Traffic Load*

Initiating our analysis, we consistently set up 100 RSUs under the client request speed of 3,000 $tps$. Fig. 11 (a) shows that as shard count increases, the traffic load per node diminishes due to the broader distribution of transactions. In GCS, nodes bear the heaviest traffic load due to the communication-intensive consensus process in PBFT [44], [45]. In contrast, TBGP and DRDST use a structured tree-based topology to optimize data flow, limiting each node to communicate only with its parent and children nodes, thus avoiding extensive pairwise messaging. Notably, DRDST exhibits the lowest node traffic load, validating the enhanced efficiency of our S-MLBT protocol.

Expanding the network to 200 RSUs with a client request speed of 3,000 $tps$, Fig. 11 (b) shows that while traffic load nearly doubles across all schemes, those using the tree-broadcasting protocol, notably DRDST, maintain significantly lower node loads. Specifically, DRDST's sharding model, which balances shard disparities, more effectively prevents hot shards, unlike GCS's geographical sharding, which can result in uneven load distribution and increased node loads.

Adjusting the client request speed to 2,000 $tps$ with 200 RSUs, Fig. 11 (c) indicates that nodes in PBFT consensus experience reduced traffic due to smaller packets with fewer transactions, while tree-broadcasting protocols face increased loads due to necessary dynamic adjustments for rate changes, especially if root node adjustments are not timely, leading to load imbalances. At lower shard counts, DRDST's frequent link reconfigurations seeking for minimum latency slightly raise node traffic over TBGP, due to the overhead of tree detachment and reconnection.

*F. Mobility of Vehicles*

When vehicles switch between RSUs across shards, efficiently managing cross-shard transactions is essential for a sharding system to maintain blockchain consensus. In this part, we specifically assess how the mobility of vehicles affects these schemes' throughput of cross-shard transactions in a 100 km² area with 100 RSUs by varying speeds of vehicles and shard counts.

Fig. 12 (a) shows that throughput increases with higher mobility in a 4-shard setup, with DRDST outperforming others due to its S-MLBT mechanism for efficient cross-shard communication. TBGP has lower throughput due to its non-latency-optimized broadcasting tree. Besides, PBFT's communication efficiency in GCS is also poor, hindering cross-shard transaction processing.

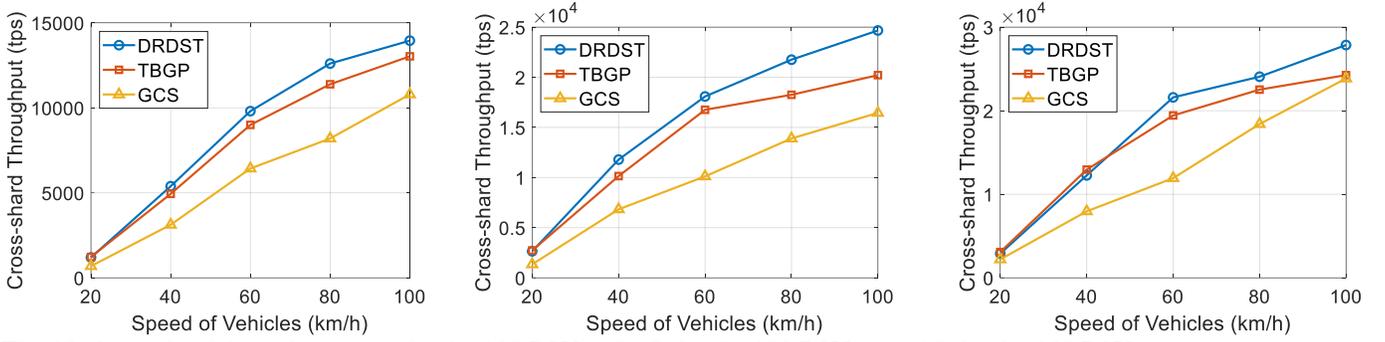

**Fig. 12.** Cross-shard throughput (a) 4 shards (100 RSUs). (b) 8 shards (100 RSUs). (c) 12 shards (100 RSUs).

When the shard count is 8, Fig. 12 (b) shows that throughput continues to rise for all schemes since vehicles encounter more shards at the same speed, leading to increased cross-shard transactions. However, TBGP's throughput grows slowly beyond 60 km/h, possibly due to FL's slow adaptation to network dynamics, leading to transaction backlogs. DRDST and GCS maintain steady growth, but sharding by RSU geography rather than performance poorly manages cross-shard transactions.

At 12 shards, Fig. 12 (c) shows only a slight further gain in throughput for all schemes, regardless of vehicle speed. Beyond 60 km/h, although DRDST maintains the highest throughput, the growth slackens, indicating the increasing difficulty in managing cross-shard transactions. The throughput gap between GCS and TBGP initially widens and then narrows. This may be because, at lower speeds, TBGP's communication efficiency is more apparent, while at higher speeds, GCS's shard backup and node scheduling improve resource allocation for consensus, enhancing performance.

*G. Ablation Study*

In this part, we perform an ablation study to empirically confirm the necessity and synergistic benefits of each component to the overall superiority of DRDST. In subsequent simulations, the system without our DAG consensus is represented as w/o DAG. It only permits one leader to initiate consensus within a shard at any given time, while the remaining nodes act solely as participants in the consensus process. The system without the S-MLBT protocol is represented as w/o S-MLBT. It employs gossip-based communication for message dissemination among nodes. The system without the robust sharding model is represented as w/o Sharding. It will utilize a random sharding method.

**Latency:** Fig. 13 (a) shows that system latency rises with more RSUs due to prolonged message propagation. The full DRDST model consistently shows the lowest latency, which moderates as RSUs increase, demonstrating effective management at scale. Removing DAG leads to escalating latency, underscoring its role in transaction ordering and cross-shard efficiency. The higher latency in the system without sharding highlights the importance of parallel processing for scalable networks. Model without S-MLBT sees a sharp latency increase, confirming the protocol's critical role in broadcast optimization. The DAG-lacking model consistently shows higher latency than the one without sharding, indicating DAG's greater contribution to network scalability and the importance of managing cross-shard transactions.

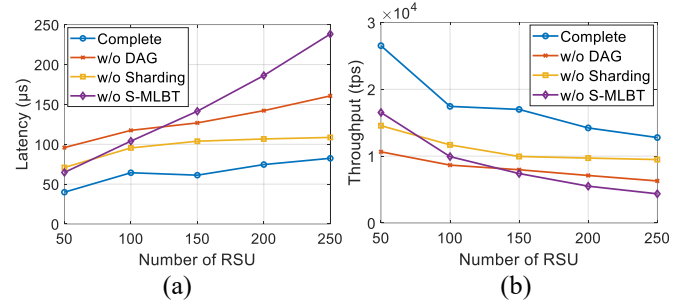

**Fig. 13.** (a) Ablation study for latency. (b) Ablation study for throughput.

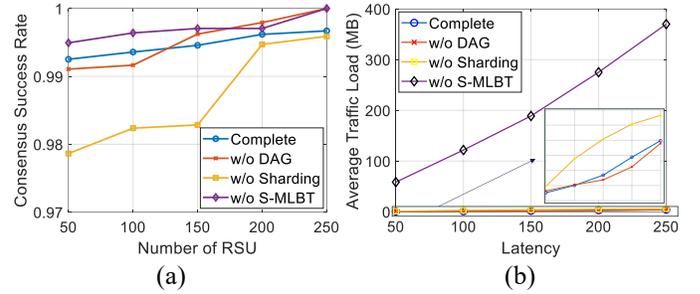

**Fig. 14.** (a) Ablation study for consensus success rate. (b) Ablation study for node traffic load.

**Throughput:** Fig. 13 (b) shows that all systems' throughput decreases as RSUs increase due to growing communication demands. The complete DRDST model retains the highest throughput, preventing hot shards and transaction backlogs effectively. The DAG-lacking model has lower throughput due to impaired high-concurrency transaction processing. Model without sharding, lacking parallel processing, also shows reduced throughput. The S-MLBT-deficient model, initially second in throughput with 50 RSUs, declines sharply to the lowest at 250 RSUs, highlighting S-MLBT's critical role in network broadcast latency and communication efficiency.

**Consensus Success Rate:** Fig. 14 (a) shows that the consensus success rate rises with increasing RSUs due to redundancy enhancing alternative consensus paths, even amidst RSU failures. The DRDST model sustains high success rates, affirming its effectiveness. The system lacking sharding suffers the lowest rates, vulnerable to malicious nodes in shards. The S-MLBT-deficient model maintains higher rates, which may be a consequence of the tree-based communication structure offering reduced latency at the expense of some robustness compared to gossip. The DAG-lacking model, despite an initial lower rate, shows an upward trend with network expansion, suggesting DAG's throughput advantages might sacrifice robustness. This highlights the importance of balancing performance and robustness in IoV consensus mechanism design.

**Node Traffic Load:** Fig. 14 (b) shows that node traffic load increases with more RSUs due to heightened communication demands for synchronization and consensus. The full DRDST model shows lower traffic loads, efficiently managing transactions and avoiding central node congestion. The DAG-lacking model exhibits decreased traffic due to reduced communication complexity from concurrent consensus initiation. Model without sharding has higher traffic loads, as sharding is key to distributing transactions and preventing overload. The S-MLBT-deficient model tops traffic loads, underscoring the inefficiency of gossip-based communication, which leads to redundant transmissions and higher load.

## VIII. CONCLUSION

In this paper, we have proposed a low-latency and robust DAG consensus to address the efficiency challenges for the IoV. We have presented a comprehensive approach that encompasses the development of a robust sharding model, the establishment of S-MLBT for optimized broadcast latency, and the implementation of an improved hashgraph protocol to efficiently manage cross-shard transactions. In this way, the impacts of vehicle mobility on blockchain consensus are well alleviated. Simulation results have demonstrated that our scheme achieves superior performance compared to other advanced schemes. Additionally, our ablation study has confirmed the critical role of each component in achieving optimal network performance, highlighting the importance of a balanced approach that considers both performance and robustness. In future work, while achieving rapid sharding, we will explore the feasibility of integrating intelligent learning algorithms suitable for IoV to further optimize sharding.